**Title**: Giant Exchange Bias in the Single-layered Ruddlesden-Popper Perovskite SrLaCo$_{0.5}$Mn$_{0.5}$O$_4$

*Ranjana R. Das[1], Priyadarshini Parida[2], A. K. Bera[3], Tapan Chatterji[4], B. R. K. Nanda[2] and P. N. Santhosh[1*]*

[1]Low Temperature Physics Laboratory, Department of Physics, Indian Institute of Technology Madras, Chennai 600036, India.

[2]Condensed Matter Theory and Computational Lab, Department of Physics, Indian Institute of Technology Madras, Chennai 600036, India.

[3]Solid State Physics Division, Bhabha Atomic Research Centre, Mumbai 400085, India.

[4]Institut Laue-Langevin, 71 Avenue des Martyrs, 38000 Grenoble, France.

E-mail: *santhosh@iitm.ac.in

**Keywords:** exchange bias, cluster spin glass, magnetic memory effect, density functional theory

Exchange bias (EB) as large as ~5.5 kOe is observed in SrLaCo$_{0.5}$Mn$_{0.5}$O$_4$ which is the highest ever found in any layered transition metal oxides including Ruddlesden-Popper series. Neutron diffraction measurement rules out long-range magnetic ordering and together with dc magnetic measurements suggest formation of short-range magnetic domains. AC magnetic susceptibility, magnetic memory effect and magnetic training effect confirm the system to be a cluster spin glass. By carrying out density functional calculations on several model configurations, we propose that EB is originated at the boundary between Mn-rich antiferromagnetic and Co-rich ferromagnetic domains at the sub-nanoscale. Reversal of magnetization axis on the Co-side alters the magnetic coupling between the interfacial Mn and Co spins which leads to EB. Our analysis infers that the presence of competing magnetic interactions is sufficient to induce exchange bias and thereby a wide range of materials exhibiting giant EB can be engineered for designing novel magnetic memory devices.

## 1. Introduction

Complex 3D perovskite oxides offer a rich materials platform for investigating emergent phenomena like ferromagnetic insulator[1], multi-ferroic[2-4], colossal magnetoresistance[5,6] that arise due to a complex web of interactions involving spin, charge, orbital and lattice degrees of freedom in the three-dimensional space [7]. In order to bring exotic quantum phenomena, a natural extension of this 3D cubic perovskite is to reduce the interaction dimensionality by either making artificial hetero superlattices or by tuning the stoichiometry so that the active layers involved in the coupling process are well separated [8-10]. While the former is sensitive to the growth condition, the latter is thermodynamically stable and can be synthesized under ambient conditions.

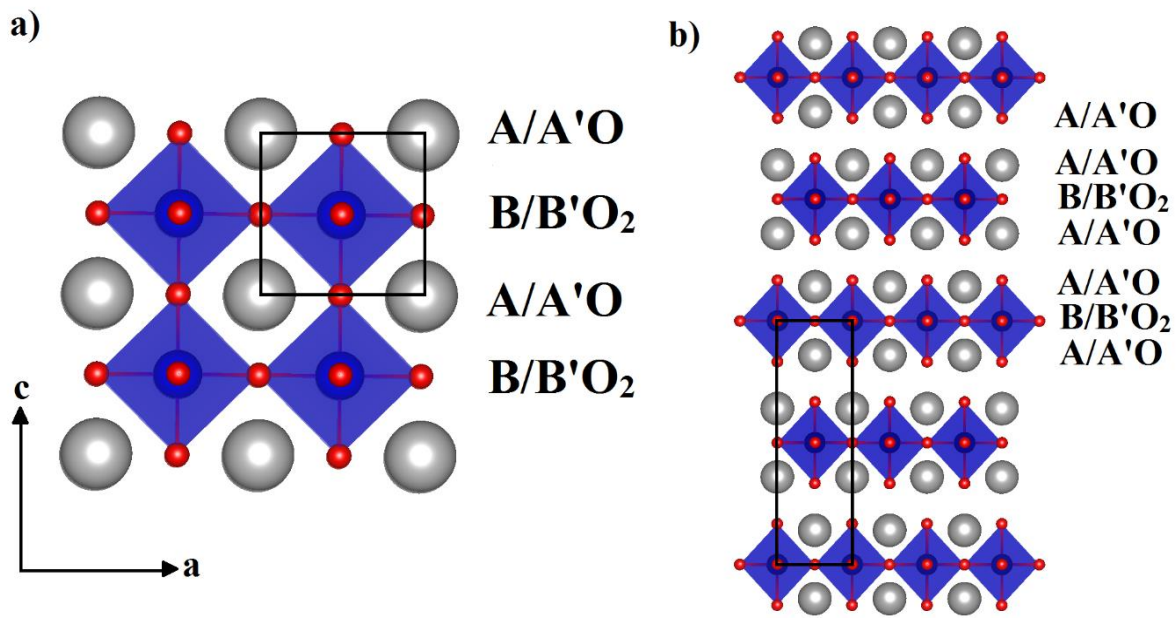

**Fig. 1. (a) Ideal perovskite structure in which the alternate stacking of A/A'O and B/B'O$_2$ layers of perovskite block is presented (square indicates one unit cell of perovskite). (b) Ruddlesden-Popper (RP) single layered in which adjacent perovskite block is replaced with a new order of the form A/A'O | A/A'O | B/B'O$_2$. The rectangle represents one unit cell of RP tetragonal lattice.**

The layered Ruddlesden-Popper (RP) series with general formula $(A)_{n+1}(B)_nO_{3n+1}$[11] are among the most preferred materials to study the physics of reduced dimensionality as they exhibit interesting properties such as superconductivity [12] [13], charge ordering (CO) or orbital ordering (OO) [14], ferroelectricity [15], and colossal magnetoresistance [16]. The crystal structure of n = 1 RP compounds, A'A(B/B')O$_4$, can be described by the periodic stacking of the layers with the order A/A'O | A/A'O | B/B'O$_2$ as shown in Fig. 1. The structure may also be described as magnetically active (A'/A)(B/B')O$_3$ perovskite blocks well separated by (A'/A)O layers along [001]. Under suitable conditions, the synthesis of these single-layer RP compounds will produce B and B' rich domains at the sub-nanoscale. The boundary separating these domains can further constrain certain long-range magnetic order, and in turn novel quantum states can be formed which may not be observed in pristine perovskites ABO$_3$. While there is a large number of literature available on single layered RP compounds with one transition metal element, very few have reported the results with two transition metal elements in a single-layered RP system [17-19]. Therefore, there is lack of evidence on emerging quantum states at the boundary of two different magnetic domains in these layered RP compounds.

In the present work, we report a novel mono-layered RP phase based layered perovskite oxide SrLaCo$_{0.5}$Mn$_{0.5}$O$_4$ (SLCMO) synthesized by sol-gel method. The dc and ac magnetic measurements and along with the powder neutron diffraction measurements point to cluster spin glass (CG) like magnetic behavior as well as a giant exchange bias (EB) effect. In fact the EB in this layered system shows fivefold increase compared to the EB reported in other layered oxide compounds such as single layer RP Sr$_{0.5}$Pr$_{1.5}$CoO$_4$ (1 kOe) [20], double layer RP Sr$_3$FeMoO$_7$ (0.2 kOe) [21], layered oxychalcogenides La$_2$O$_3$Mn$_2$Se$_2$ (0.5 kOe) [22] and double perovskites SrLaCoMnO$_6$ (0.3 kOe) [23]. The density functional calculations further confirm the existence of giant EB in this compound, and it attributed to the anisotropic magnetic coupling among the Mn and Co spins.

## 2. Experimental Section

*Synthesis details*:

Despite remarkable electronic and magnetic properties, the synthesis of phase pure stochiometric single layer RP system is quite challenging. The probability of getting a perovskite phase was often detected in the polycrystalline sample, which was synthesized by solid-state synthesis methods [24]. Thus a citrate gel technique is used to prepare a polycrystalline sample of SrLaCo$_{0.5}$Mn$_{0.5}$O$_4$ (SLCMO). La$_2$O$_3$, Mn(CH$_3$COO)$_2$.4H$_2$O, Sr(NO$_3$)$_2$ and Co(NO$_3$)$_2$.6H$_2$O in stoichiometric amounts were first dissolved in dilute nitric acid (HNO$_3$), and then an excess of citric acid and ethylene glycol (CH$_2$OH)$_2$ was added. La$_2$O$_3$ was preheated at 1273 K before adding in dilute nitric acid. The dissolved solution was heated on a hot plate resulting in the formation of a gel. The gel was dried at 523 K and then heated to 973 K for 12 h to remove the organic components and to decompose the nitrates. SLCMO ceramic was subsequently sintered at 1573 K for 36 h in the air with intermittent grindings for homogeneity.

*Characterization details*:

Laboratory X-ray powder diffraction (XRPD) measurements were conducted at room temperature on a PANalytical X′pert using a Cu Kα radiation source (λ = 1.54056 Å). The neutron powder diffraction (NPD) patterns were recorded in zero field by using a high-resolution powder diffractometer SPODI at the research reactor FRM-II (Garching, Germany) with monochromatic neutrons of 1.5481(1) Å over the 2θ range of 6 – 150˚ with a step size of 0.05˚. X-ray photoelectron spectra (XPS) of SLCMO were recorded by an instrument with Mg-Kα as the X-ray source, and a PHOIBOS 100MCD analyzer (SPECS) operated under ultra-high vacuum (10$^{-9}$mbar). The XPS spectra were fitted by the CasaXPS software (Casa Software Ltd) using Gaussian-Lorentzian peak functions and Shirley background subtraction. Dc and ac magnetic susceptibility measurements were carried out in a commercial superconducting quantum interference device (VSM-SQUID) magnetometer 70 kOe (MPMS). Dc magnetization curves in zero-field cooled (ZFC) and field cooled (FC) cycles were performed at several magnetic fields. Both the ZFC and FC magnetization versus temperature (M-T) curves were measured during the warming process. Isothermal magnetization, M (H), hysteresis loops were measured under ZFC and FC conditions in the range −70 kOe ≤ $\mu_0H$ ≤ 70 kOe at different temperatures. For the ZFC case, the samples were first cooled from 300 K down to the temperature of measurement under zero magnetic field, and five-quadrant M(H) measurements were performed starting at H = 0. For FC analysis, the samples were cooled each time under an applied magnetic field from 300 K down to measurement temperature.

*Computational Details:*

Pseudopotential based density functional calculations are carried out using the plane waves as basis sets as implemented in Quantum ESPRESSO [25]. We have used the generalized gradient approximation (GGA) exchange correlation functional, given by Perdew, Burke and Ernzerhof, incorporated with the Hubbard U correction to consider the strong correlation effect. We have taken U = 5 *eV* in our calculations. In all our calculations ultrasoft type pseudopotentials are considered. The kinetic energy cut-off for the plane waves and charge densities are taken as 30 Ry and 250 Ry, respectively. The k-mesh of 3×9×4 with 30 irreducible k-points are used in the calculations.

## 3. Experimental results

### 3.1. Crystal structure

#### 3.1.1. X-ray and Neutron diffraction

The crystal structure of SLCMO is examined through X-ray powder diffraction (XRPD) and neutron powder diffraction (NPD) measurements. NPD measurements are performed at various temperatures ranging from 300 K to 4 K. The Rietveld refinement of NPD patterns collected at 300 K, and 4 K are shown in Fig. 2 and Rietveld refinement at 50 and 100 K can be seen in the supplementary file. The data indicate that the compound crystallizes in a single-layered RP phase with the body-centered tetragonal lattice having space group *I4/mmm* No. (139) and the schematic of this structure is presented in Fig. 1(b) (where A/A′= Sr/La as silver balls, B/B′ = Co/Mn as blue balls and O as red balls). The room temperature lattice parameters ′a′ and ′c′ are found to be 3.8420 (1) Å and 12.6211 (1) Å respectively from the NPD. The lattice parameters *a* and *c* change by 0.23% and 0.17% respectively over the temperature range 4 - 300 K. Details of crystal structure parameters obtain from XRPD and NPD are provided in the supplementary material. From the nuclear structure refinement, we conclude that SLCMO is stoichiometric. We observed a large difference between the equatorial and apical bonds of Mn/Co-$O_6$ octahedra with bond length 1.92 Å and 2.09 Å respectively, and the difference in bond length is close to 0.17 Å. We may note that Jahn-Teller active $LaSrMnO_4$ has a similar difference in the bond length [26]. The obtained bond lengths (Table S3 supplementary) from the NPD refinement signatures the compression of equatorial and elongation of apical bonds indicating the presence of Jahn Teller active $Mn^{3+}$ ion ($t_{2g}^3 e_g^1$).

In order to get an insight into the oxidation states of the ions, we have performed bond-valence-sum (BVS) calculations from the refined atomic positions. The derived BVS values reveal a mixed valence state of both $Co^{2+/3+}$ and $Mn^{3+/4+}$ ions.

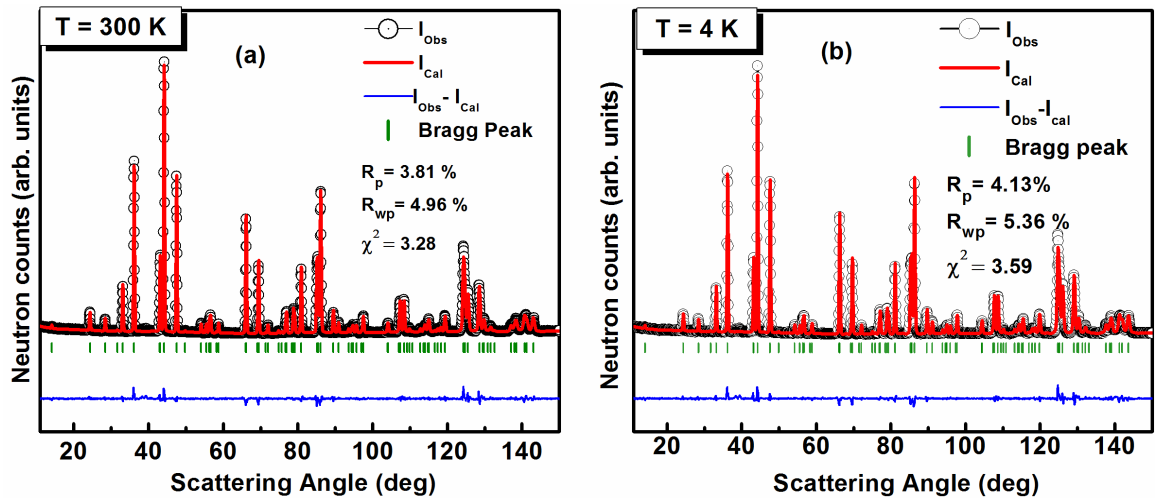

**Fig. 2 Neutron powder diffraction pattern of $SrLaCo_{0.5}Mn_{0.5}O_4$ measured at (a) 300 K and (b) 4 K. We excluded the *2θ* region 38 – 40° of the neutron diffraction patterns during Rietveld refinements, as it consists of two weak peaks from the sample holder.**

Temperature evolution of NPD shows that there is no structural change in the measured temperature range. Furthermore, no additional magnetic peak or increase in intensity is observed which confirms the lack of long-range magnetic ordering in the extended temperature range [4 – 300 K].

*3.1.2 X-ray photoelectron studies (XPS)*

XPS measurements provide information on the surface composition and therefore we performed Mn-2p and Co-2p core level XPS to assign the atomic oxidation states of Mn and Co in SLCMO compound. The spin-orbit splitting of Mn 2p peaks corresponds to Mn $2p_{3/2}$ and Mn $2p_{1/2}$ which are located at 641.97 eV and 653.38 eV, respectively (Inset: Fig. 3 (a)) whereas that of Co 2p peaks (Inset: Fig. 3 (b)) are found at 779.93 eV (Co $2p_{3/2}$) and 795.33 eV (Co $2p_{1/2}$). The oxidation state of Mn/Co ions was determined by the curve fitting the corresponding 2p spectral peaks. The experimental peak shape for Mn $2p_{3/2}$ and Co $2p_{3/2}$ was modeled by employing double peaks (Gaussian-Lorentzian) patterns and shown in Fig. 3(a) and (b) respectively. Both Mn $2p_{3/2}$ and Co $2p_{3/2}$ spectra show perfect fitting for mixed valence states. The two binding energy values obtained for Mn $2p_{3/2}$ at 641.17 eV and 643.87 eV matched well with the reported values for $Mn^{3+}$ and $Mn^{4+}$, respectively [27,28]. Similarly, Co $2p_{3/2}$ spectra matches with the reported binding energy values of 3+ and 2+ state at 779.8 *eV* and 780.2 *eV* respectively [27,28]. Peak fitting corresponding to Mn $2p_{1/2}$ and Co $2p_{1/2}$ spectra are given in the supplementary files which again confirms mixed valance state. From the XPS fitting, we have estimated the percentage of $Mn^{3+}$/ $Mn^{4+}$ to be 67/33 % and that of $Co^{3+}$/$Co^{2+}$ to be 65/35 %. Hence, the predominant oxidation states in SLCMO is confirmed to be $Mn^{3+}$ and $Co^{3+}$ which also corroborates the charge states obtained from the neutron diffraction data. These observations lend strong support to the counter part of DFT calculations.

**3.2. Magnetic properties:**

**3.2.1. dc Susceptibility**

Fig. 3(c) shows the temperature dependence of the dc magnetic susceptibilities χ(T) of SLCMO in the zero field cooled (ZFC) and field cooled (FC) conditions under applied fields of 100, 200 and 1000 Oe. ZFC curve shows two broad peaks (humps) at ~ 150 K and ~ 50 K, whereas, the FC curve shows a sharp rise at ~ 140 K followed by an anomaly below ~ 50 K (slope change) at low temperatures. The temperature dependent derivative curves (dM$_{ZFC}$/dT) (inset of Fig. 3c) reveal these two transitions at T$_{C1}$~ 150 K and T$_{C2}$~ 50 K. The susceptibility data under an applied field of H =100 Oe (ZFC) in the temperature range 180-300 K were fitted with the Curie-Weiss (CW) law, i.e., $χ^{-1}$ (T)= C/(T-$θ_p$), (Fig. 3(d)) where C and θ$_P$ are Curie constant and CW temperature respectively. The fit provided positive values of CW constant [$θ_P$ = 99 K] suggesting the presence of dominant ferromagnetic interactions. The effective paramagnetic moment calculated from the Curie constant of C = 6.42 emu.K.mol$^{-1}$Oe$^{-1}$ is μ$_{eff}$= 4.86 μ$_B$ / f.u. ( theoretically estimated value  4.54 μ$_B$ (μ$_{the}$ = √(4S(S + 1))μ$_B$) on considering high spin (HS) state of Mn$^{3+}$/Mn$^{4+}$ [67/33 %]  and Co$^{3+}$/Co$^{2+}$ [65/35 %] from the XPS are taken into account). In general, a strong ferromagnetic compound with long range magnetic interaction exhibits CW temperature $θ_P$ to be equal or greater than Curie temperature T$_C$. However, in disordered systems with randomly distributed magnetic ions provide competing magnetic interactions and spin frustrations resulting in $θ_P$ < T$_C$ [29]. The bifurcation between ZFC and FC arms and the lower temperature transition at 50 K indicate that there might exist spin glass or glassy like magnetic interaction in addition to the FM interactions, which will be clarified in later sections.

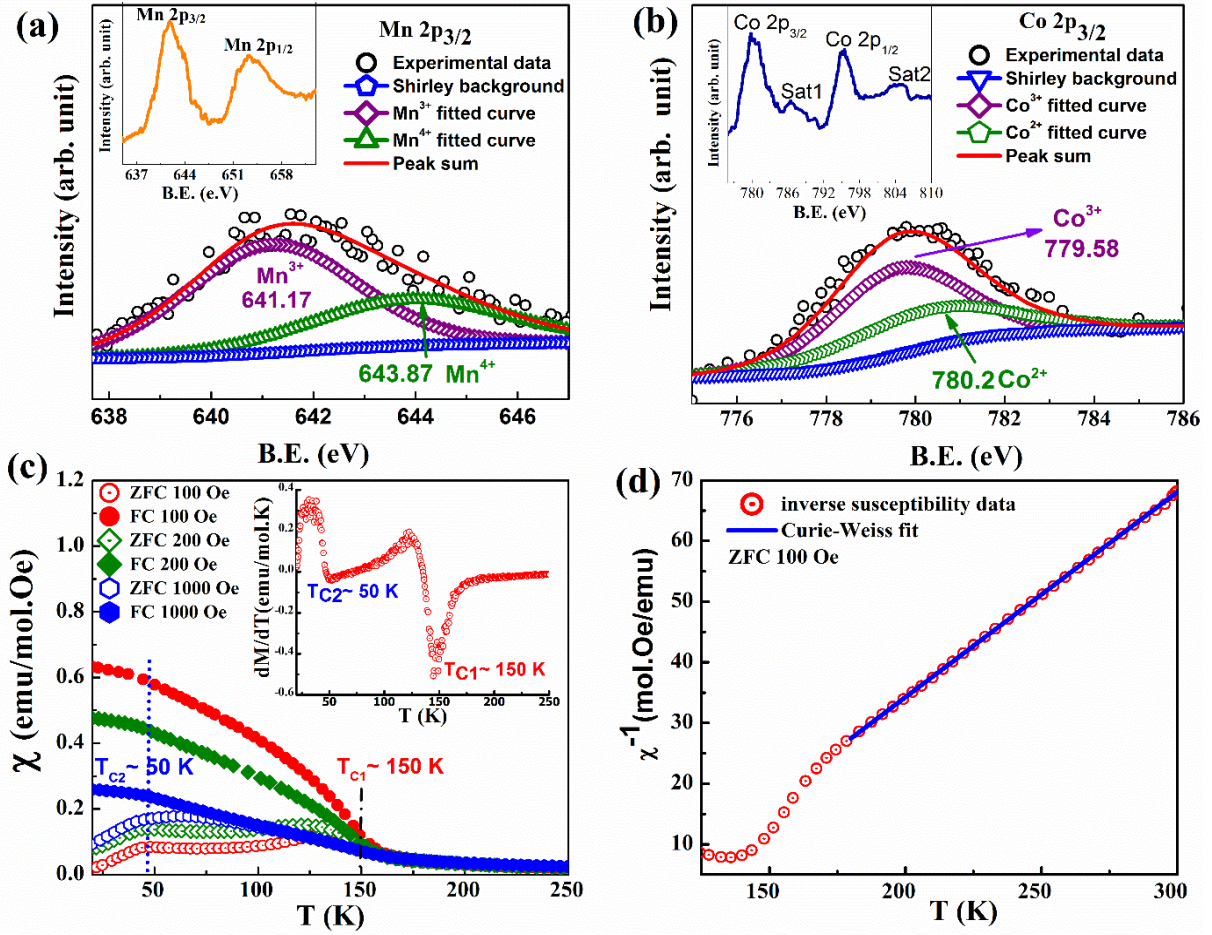

**Fig. 3** X-ray Photoemission spectroscopy of **(a)** Mn 2p$_{3/2}$ (inset shows the Mn 2p core level spectrum) and **(b)** Co 2p$_{3/2}$ along with fitted curves (inset shows the Co 2p core level spectrum). **(c)** Magnetic susceptibility χ measured under 100, 200 and 1000 Oe with ZFC and FC protocols. The dotted vertical lines indicate two transitions at T$_{C1}$ = 150 K and T$_{C2}$ = 50 K respectively. (Inset shows the derivative of magnetization vs. temperature curve (ZFC) measured under 100 Oe). **(d)** The inverse magnetic susceptibility (χ$^{-1}$) vs. temperature curve, for ZFC at 100 Oe.

### 3.2.2. Memory effect

The dc thermomagnetic analysis of SLCMO suggests glassy magnetic transition at lower temperatures. Magnetic memory effect is an experimental fingerprint of glassy magnetic material. We have performed the same to figure out the glassy magnetic features in SLCMO. We have carried out a detailed magnetic memory analysis using both ZFC and FC experimental protocols [30]. The ZFC memory experiments were carried out by conventional procedure with cooling the sample without field at a constant rate of 2 *K/min* with an intermediate halt at 40 K for a duration of 1h. The magnetization data with respect to ZFC memory were recorded during the warming cycle without any halt under an applied dc field of 100 Oe. Fig. 4(a) depicts the temperature dependence of ZFC reference magnetization $M_{ZFCW}^{ref}$ and ZFC memory magnetization $M_{ZFCW}^{mem}$ along with the differential curve. A sharp memory dip in the differential curve at 40 K shows the clear time evolution of magnetization at the stopping temperature and confirms the glassy dynamics in SLCMO below $T_{C2}$. For FC memory, first the sample was cooled from 300 to 5 K at a constant rate of 2 *K/min* under a cooling field of 100 Oe with recurring stops at 100, 40 and 10 K for a duration of 1h at each stops. During each stop the field was set to zero to let the magnetization relax downward. After each stop and wait period (where we see a sharp step during each halted temperature in figure 4 (b) blue solid line with open diamond symbol with dot), the 100 Oe field is reapplied and cooling is further resumed. This cooling procedure produces a sharp jumps in the M(T) curve at the halt temperatures. The data thus obtained is considered as $M_{FCC}^{stop}$. After reaching the base temperature 5 K, the sample temperature is raised continuously at the 2 *K/min* rate in a constant 100 Oe field and the magnetization is recorded again. This curve is called $M_{FCW}^{mem}$ curve. Fig. 4(b) represents the FC memory plots where $M_{FCC}^{stop}$ curve exhibits sharp jumps at halt temperatures (100, 40 and 10 K) while $M_{FCW}^{mem}$ curve exhibits two clear upturns near the stopping temperatures at 10 and 40 K which is below $T_{C2}$. However, $M_{FCW}^{mem}$ curve portraits a continuous curve without any noticeable change at the stopping temperature of 100 K (see Fig. 4(b) solid red line) which is above $T_{C2}$. Thus, clear magnetic memory effects is observed at temperatures 10 and 40 K confirming that the present compound remembers its previous history of zero field relaxation only below $T_{C2}$ due to the slow dynamics of frozen spins in this region. Hence, it is confirmed from both ZFC and FC memory experiments that the low temperature transition exhibited by SLCMO at $T_{C2}$ is a glassy magnetic transition and the high temperature transition at $T_{C1}$ is a ferromagnetic transition.

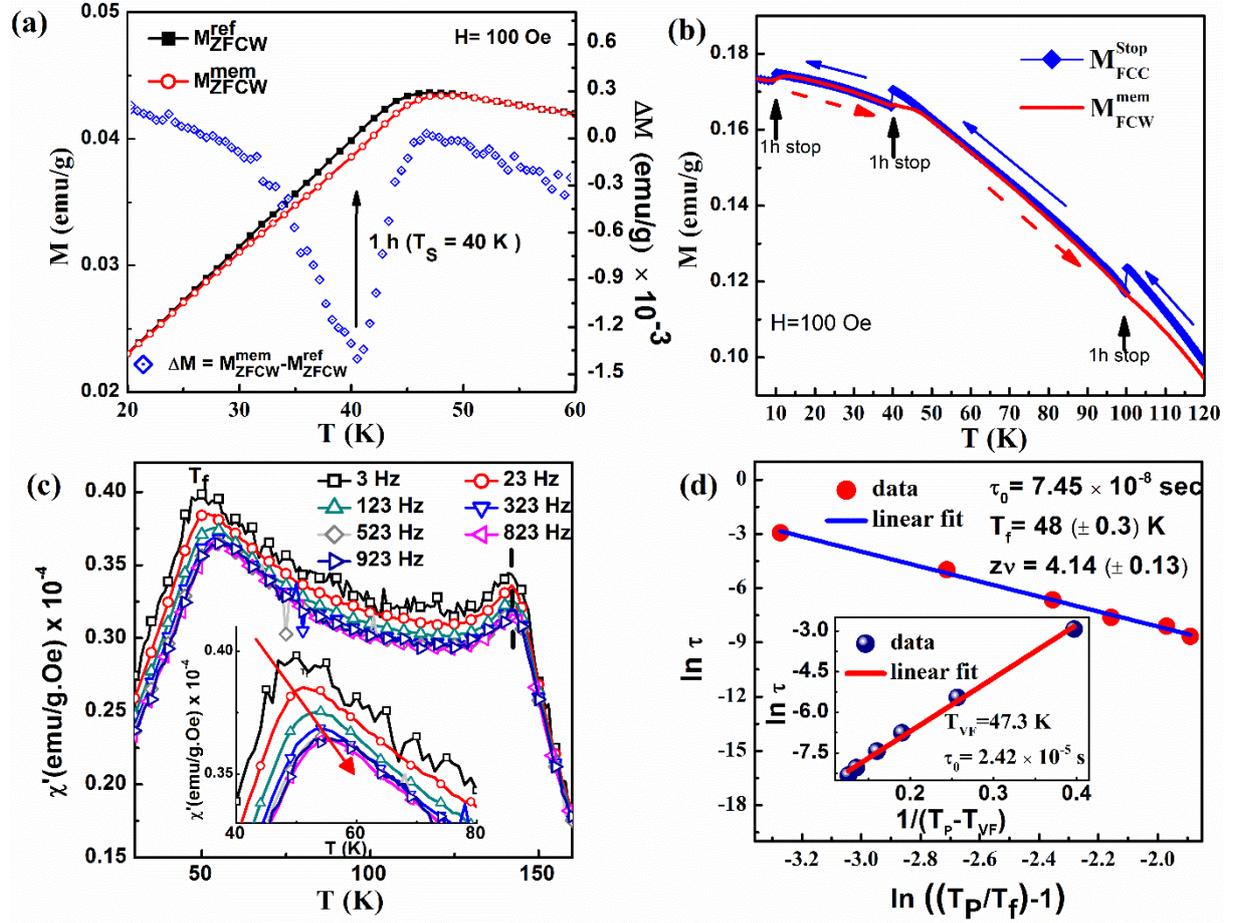

**Fig. 4. (a) ZFC memory effect measured after halting for 1 h at 40 K while cooling from 300 K to 5 K and then measurement was done in the warming cycle under an applied field of 100 Oe. The difference curve $\Delta M = M_{ZFCW}^{ref} - M_{ZFCW}^{ref}$ as a function of temperature. For the reference curve, the sample was cooled to 5 K from 300 K without a halt and measurement was done in the warming cycle. (b) FC memory effect experiment: Intermittent-stop cooling magnetization $M_{FCC}^{stop}$ at 100, 40 and 10 K while cooling from 300 K to 5 K (marked as solid blue arrow) and the red solid line is the continuous warming memory curve $M_{FCW}^{mem}$ (marked as dashed red arrow). (c) Temperature dependence of the real part of ac susceptibility (χ′) measured under different frequencies with ac magnetic field of 2 Oe. (inset the plot of $\ln \tau$ vs. $\ln \left( \frac{T_P}{T_f} - 1 \right)$ (red solid circles) and the best fit to Eq.2 (solid blue**

**line) shown and inset shows the zoomed portion of T$_{C2}$. (d) $ln\,\tau$ vs. $\frac{1}{(T_P - T_{VF})}$ plot for cluster glass transition and the solid line is the linear fit for Vogel-Fulcher law (Eq. 1)**

### 3.2.3. ac Susceptibility:

To confirm the true nature of magnetic ordering in this compound, we have measured the temperature-dependent ac susceptibility ($\chi_{ac}$), under an ac field of 2 Oe within the frequency (f) range of 3 Hz to 923 Hz. In agreement with the dc susceptibility data, the temperature dependence of the real part of the ac-susceptibility curves ($\chi'$) show two pronounced peaks at around ∼ 150 and 50 K as shown in Fig. 4(c). The frequency variation study reveals that the position of the high-temperature peak (∼150 K) does not shift, and all the curves at various frequencies merge well above ∼ 150 K confirming paramagnetic to ferromagnetic phase transition at T$_{C1}$. Whereas, the peak at ∼ 50 K has a pronounced frequency dependence shift, i.e., with increasing frequencies, the peak shifts towards high temperature from 49.82 K at 3 Hz to 55.25 K at 923 Hz, with a decrease in peak amplitude as shown in the inset of Fig. 4(c). Such behavior is a characteristic feature of the spin glass (SG) and/or disordered magnetic systems [31]. In order to differentiate between the SG and CG behavior, we have carried out the analysis of the susceptibility data at T$_{C2}$ using the Mydosh parameter, Vogel-Fulcher (VF) relation, and slow dynamics models [32]. First, the Mydosh parameter ($K = \frac{\Delta T_P}{T_P \Delta(log_{10} f)}$) which is empirically known as the relative frequency shift in the peak temperature $T_P$ of ($\chi'$) per decade of frequency [31], is found to be 0.04 (±1) and this value is comparable to the values reported for other CG systems ($K \leq 0.08$) [33,34].

Secondly, we used Vogel–Fulcher (VF) relation to understand the characteristic relaxation time, which diverges at the freezing temperature T$_{VF}$

$$\tau = \tau_0 exp\left[\frac{-E_a}{k_B(T_P - T_{VF})}\right], \text{-------------(1)}$$

where $\tau_0$ represents the characteristic relaxation time of the clusters, $E_a$ is the activation energy, and $T_{VF}$ is the VF freezing temperature which provides inter-clusters interaction strength. The expression is valid for peak temperature ($T_P$) greater than $T_{VF}$ [32]. The linear fitted $ln\,\tau$ versus $1/(T_P - T_{VF})$ curve, obtained by the Souletie and Tholence method [35] with $T_{VF}$ = 47.3 K, is shown in inset of Fig. 4(d). The curve yields $E_a/k_B$ = 19.65 (4) K and $\tau_0$ = 2.42 x $10^{-5}$ s, which falls within the range of characteristic relaxation times for CG [34].

Finally, we have fitted experimental data (Fig 4(d)) by the conventional critical slowing down dynamics model,[31]

$$\tau = \tau^* \left(\frac{T_P - T_f}{T_f}\right)^{-z\nu}, \quad\quad\quad\quad\quad\quad\quad (2)$$

where $\tau$ is the relaxation time corresponding to the measured frequency, $\tau^*$ is the microscopic relaxation time, $z\nu$ is the dynamic critical exponents, and $T_f$ is the static finite freezing temperature for $f \to 0$ Hz. We observed that our data well fitted to Eq. (2) as shown in Fig 4(d); the best fit yields $\tau^* \approx 7.45 \times 10^{-8}$ s and $z\nu \approx 4.14$ ($\pm 0.13$), with $T_f \approx 48$ K. The critical exponent falls inside the typical range for glassy magnetic systems ($z\nu \sim 4$ -12), and observed fitted values are close to those reported for CG ($\tau^* \approx 10^{-8}$ s and $z\nu \approx 6$) [36]. The observed lengthier relaxation time obtained from both VF, and critical slowing down relaxation exemplars in conjunction with the calculated Mydosh parameter value confirm the CG behavior of the sample below $T_f$ at 48 K.

### *3.2.4. Isothermal magnetization and Exchange bias*

Fig. 5(a) shows the *M - H* plot for SLCMO sample measured at 5, 50, 100, 150 and 250 K. The linear isothermal magnetization curve at 250 K indicated the paramagnetic phase of the sample. For T = 150 K and below, the *M-H* curve diverged from the linearity, and the hysteresis loop becomes more prominent below $T_{C1}$. An enlarged view of hysteresis loops at 50, and 100 K are given in the inset of Fig. 5 (a). The absence of long range ordering between the magnetic cations (which is confirmed from neutron diffractions) along with the presence of frozen spins at lower temperatures confirms the coexistence of ferromagnetic–antiferromagnetic phases in SLCMO. Therefore, the transition below 150 K is a weak ferromagnetic transition resulting with low coercivity below 150 K down to $T_{C2}$ (100 K: $H_C$ ~ 200 Oe and 50 K: $H_C$ ~ 600 Oe). However, in SLCMO an enhance coercivity is observed below CG ($T_{C2}$) transition and it reaches to a maximum value of about ~ 7500 Oe at 5 K. As the system enters into CG region the coercivity is enhanced due to spin freezing phenomena. When the magnetic moments freeze below $T < T_{C2}$, the spins are trapped in the increased free energy barriers between multiple energy states. So, in an applied field the magnetization direction is flipped, and thus the coercivity is enhanced in order to overcome the increased free energy barrier [37]. The estimated magnetization measured at 5 K and high field of 70 kOe is 0.8 $\mu_B$ / f.u. which is smaller than the expected spin only saturation magnetization value of 3.66 $\mu_B$ / f.u. (this value is calculated by considering the spin states ratio obtained from XPS with Mn and Co ions as high spin states). This massive reduction in observed magnetic moments at 5 K strongly implies the presence of competing AF interactions inside the multiple ferromagnetic islands.

Exchange bias is a phenomena formed due to the magnetic anisotropy created at the interface of AFM and FM phases [38,39]. There are several reports regarding the presence of Exchange Bias (EB) in magnetic oxides with competing magnetic interactions [40-42]. Since the present system possesses different magnetic interactions lead to CG and FM transitions, we have comprehensively investigated the EB effect in SLCMO. The magnetic hysteresis measurements at 5 K was performed in ZFC as well as FC cycles at different values of applied cooling field (CF) and temperatures to understand EB for this system in detail. In FC mode the sample was cooled under a magnetic field of 50 kOe (other than specified) from 300 K to the measurement temperature.

In earlier reports [43,44], it has been noticed that the incorrect optimization of maximum measuring fields can lead to the existence of a minor loop effect. To rule out the minor loop effects and to ensure the observation of a genuine EB-shift, we have considered the following point: Anisotropy field ($H_A$) of the system should be less than the optimal maximum applied field ($H_{max}$).

Now to obtain the anisotropy field, we have used the law of approach for saturation magnetization equation on the initial magnetization virgin curve at 5 K [45] i.e.

$$M = M_S * \left(1 - \frac{a_1}{H} - \frac{a_2}{H^2}\right) + \chi H \quad\text{-----------------------}(3)$$

Where $a_1$ and $a_2$ are the free parameters, $M_s$ is the saturation magnetization, and $\chi$ is the high-field susceptibility. According to Andreev et al.[45] the first term in Eq. (3) is related to a local anisotropy which originates from the structural defects and nonmagnetic inclusions of local magnetization. While the second term corresponds to the rotation of magnetization against the magnetocrystalline energy. In case of a high anisotropic ferromagnet compounds $a_1 \ll a_2$ ($a_2 = \frac{4K_1^2}{15M_S^2}$), where constant $a_2$ provides the estimation of magnetocrystalline anisotropy.

A rough estimation of the anisotropy field $H_A = \frac{2K_1}{M_s}$ was obtained by using Eq. (3) (where $K_1$ is an anisotropy constant). From 5 K *M-H* data, the anisotropy field of ∼20 kOe obtained by using Eq. (3), which is comparable to the value reported for Fe-doped $LaMnO_3$ [46]. The values of $H_{max}$ were considered much higher than $H_A \sim 20$ kOe for the major hysterisis loop tracing. Along with to rule out the minor loop presence, we have recorded the hysteresis loop under positive and negative CF of 50 kOe. The hysteresis loops are found to be shifted towards the opposite direction to the applied CF, which is a signature of conventional EB presence in this compound [Fig. 5(b) shows zoomed versions of the hysteresis loops between ±50 kOe for the sake of clarity].

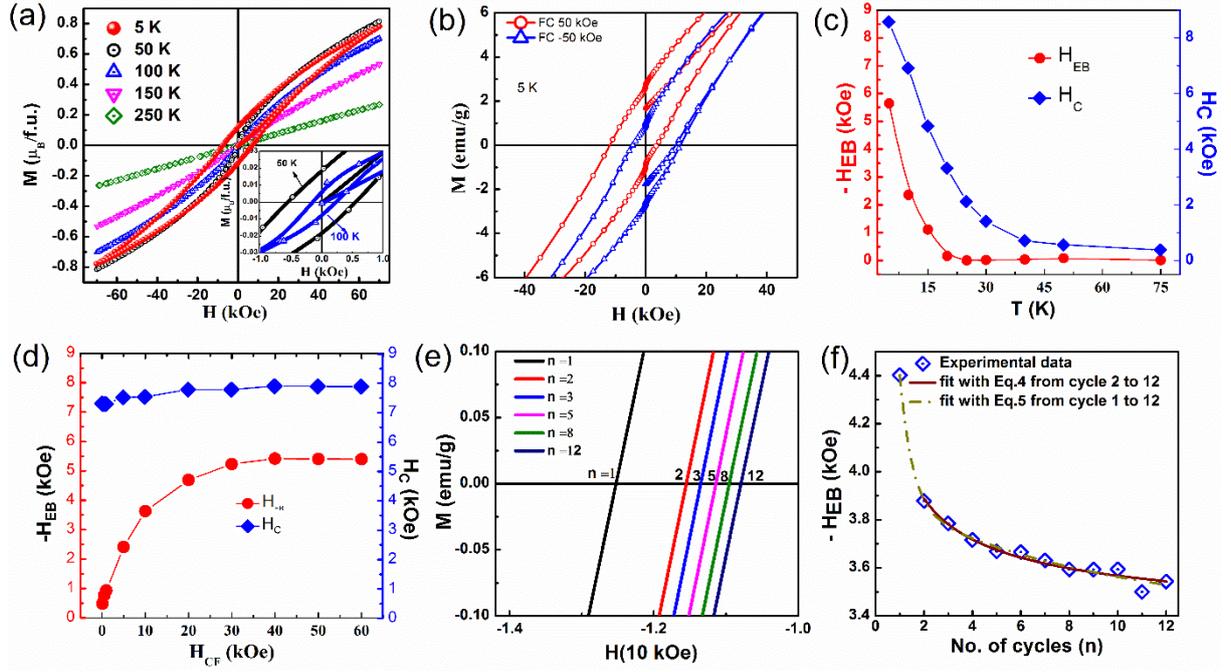

**Fig. 5** (a) M - H loops at different temperatures in the range ±70 kOe (Inset shows the zoom of the interior region in the curve at 50 and 100 K). (b) The M-H hysteresis loops measured at 5 K after cooling the sample from 300 K under 50 kOe field. The measurement range is between ± 70 kOe. For clarity, only the data between ± 50 kOe are shown. (c) $H_{EB}$ and $H_C$ as a function of cooling fields at 5 K. (d) Temperature-dependent $H_{EB}$ and $H_C$ in cooling field 50 kOe. (e) Magnetization curves measured at 5 K after field cooling ($H_{FC}$ = 50 kOe) with 12 continuous loops and zoomed the left side of the hysteresis curves around M=0, apparently, notice the significant difference from n = 1 to n = 2 than any other consecutive loops. f) Training effect of FC exchange bias field ($H_{EB}$) vs. no of hysteresis at 5 K. The lines are fitted curves by two different conditions.

Traditionally, EB and coercivity values are obtained as $H_{EB} = \left(\frac{H_{c1}+H_{c2}}{2}\right)$ and $H_C = \left(\frac{|H_{c1}|+|H_{c2}|}{2}\right)$, where $H_{c1}$ and $H_{c2}$ are left and right coercive fields respectively. The hysteresis curves in a cooling field of 50 kOe measured at different temperatures from 5 to 75 K in the field range ±70 kOe and the value of $H_{EB}$ and $H_C$ are plotted as a function of temperatures [Fig. 5(c)]. The $H_{EB}$ vs. temperature curve shows an exponential decay behavior. One can see that the EB appears below $T_f = 48\ K$ of the CG state. At 5 K, the displacement of FC loop becomes much more prominent with an $H_{EB}$ = 5.5 kOe at a CF of 50 kOe, which is ten times larger than that of the double perovskite compound LaSrCoMnO$_6$ involving same magnetic ions (measured in the same condition, i.e., at 5 K with CF 50 kOe) [23]. Also, we have performed the cooling field from 0.01 kOe to 60 kOe dependence of the EB (Fig. 5(d)) at 5 K with a maximum field range of 70 kOe. We observe a sharp increase in both $H_{EB}$ and $H_c$ with increasing CF up to 40 kOe with a giant EB of ~5.5 kOe followed by a more gradual saturation of this effect at higher CFs up to 60 kOe. When cooling the specimen to $T < T_f$, in the presence of a magnetic field, the CG spins next to the FM/AFM spins arrange along specific direction due to the exchange interaction at the frustrating interface. As a result, there will be strong pinning between frozen FM and AFM island of spin clusters and the FM spins producing EB effect along the CF direction.

Training effects are complementary characteristics of EB phenomena and occurring by the non-equilibrium nature of the spin structures in the pinning layer [47]. While cycling the system through several consecutive hysteresis loops, it is manifested as the gradual decrease in $H_{EB}$ and showed a clear indication of rearrangements in the pinning layer spin structure towards an equilibrium configuration supporting Binek′s proposition on EB effect in antiferromagnetic-ferromagnetic heterostructures. Later Mishra et al. [48] proposed that local spins of AFM side of the interface be affected from both the components of frozen and rotating spins by the FM magnetization reversal. In this view, a series of twelve continuous hysteresis loops were measured at 5 K over ± 70 kOe under a CF of 50 kOe as shown in Fig. 5(e), which illustrates a close view of the left side of hysteresis curves around $M = 0$ axis. The $H_{EB}$ values obtained from each M (H) are plotted with the number of hysteresis cycles (n) in Fig. 5(f). A monotonic decrease of the EB effect is observed with continuous loops measurement. The following power-law function which can describe the reduction of $H_{EB}$ as a function of $n$ in terms of energy dissipation of the AFM regions at the pinning interfaces:

$$H_{EB} - H_{EB}^{\infty} \propto n^{-\frac{1}{2}}, \text{--------------------(4)}$$

where $n$ is the loop index number and $H_{EB}^{\infty}$ the value at $n = \infty$ which is 3.188 kOe for the present sample. The Eq. (4) holds only for loops from $n \geq 2$, and cannot explain the sheer relaxation between the first and second loops as shown in Fig. 5(f). According to Mishra et al. [48], interfacial spin frustration can occur at the magnetically disordered FM/AFM interface due to AFM magnetic anisotropy. This magnetic anisotropy is contributed from two different types of AFM spins after field cooling: specifically, frozen and rotatable AFM spins [48]. As this compound exhibit EB below the CG transitions with disordered FM/AFM phases, it is appropriate to use the model (Eq. 5) proposed by them for fitting the training effect. The equation which satisfies the condition is expressed as

$$H_{EB}^{n} = H_{EB}^{\infty} + A_f \left(\frac{-n}{P_f}\right) + A_r \left(\frac{-n}{P_r}\right), \text{--------------------(5)}$$

where f and r denote the frozen and rotatable AFM spin components at the pinning interface. The parameters, '$A$' have dimensions of magnetic field (Oe), whereas, parameters, '$P$' are dimensionless quantities identified with relaxation. As can be seen from Fig. 5(f), the FC EB training effect data fit well with Eq. 5 in comparison to Eq. 4. The parameters obtained from the fit to the $H_{EB}$ data are $H_{EB}^{\infty}$ = 3.4 kOe, $A_f$ = 3.8 kOe, $P_f$ = 0.51, $A_r$ = 0.505 kOe, and $P_r$ = 8.71 which suggest that the rotatable components are relaxed 17 times faster than the frozen spin component at the interface in the presence of cooling field of 50 kOe. Similar phenomena has been obseved in a spin glass system of $La_{1.5}Ca_{0.5}CoIrO_6$ [40].

## 4. Estimation of exchange bias from DFT calculations

The experimental results conclusively show the CG magnetic structure and a giant exchange bias for SLCMO RP structure. The CG state of the system suggests coexistence of AFM and FM rich domains and the EB occurs at the boundary between these domains. Experimentally it has been reported that the $Sr_{2-x}La_xCo/MnO_4$ has a rich magnetic phase diagram depending on the La and Sr concentration [49-53]. Our electronic structure calculations along with the reported literature confirm that pristine $LaSrCoO_4$ and $LaSrMnO_4$ have FM [54] and AFM [49,55-58] states, respectively. However, $LaSrCoO_4$ is also reported as a spin glass in a recent article [59]. The earlier discussions on the magnetic measurements have revealed the coexistence of both AFM and FM phases at low temperature (< 50 K). Further, short-range magnetic ordering are inferred from the combined study of neutron diffraction, dc- and ac-susceptibility measurements. However, the XRPD/NPD do not indicate any Co and Mn-rich segregated phases. Therefore, the coexistence of the FM and AFM ordering can be explained provided there is a sub-nano scale (two to three unit cell length) Co and Mn-rich domains. The primary arrangements of such FM and AFM domains are shown in Fig. (6). The secondary arrangements can be obtained through vector combination of these primary arrangements. Using DFT calculations, the strength of exchange bias is estimated by spin-flipping mechanism [60].

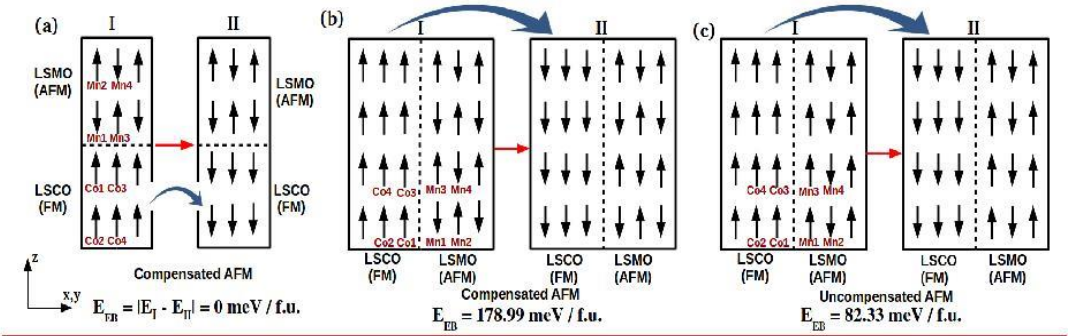
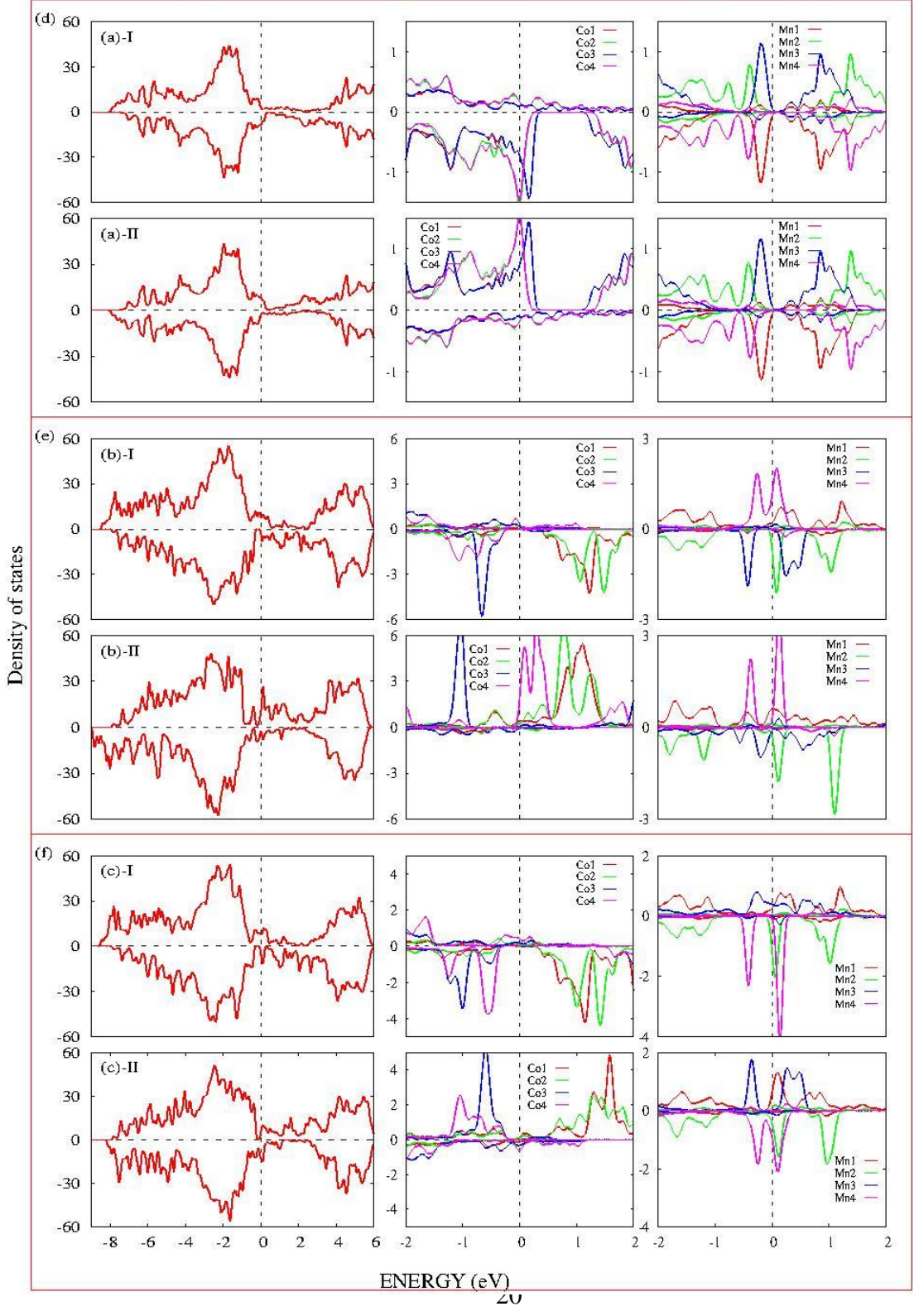

**Fig. 6 Schematics for three different possible interfaces between FM LaSrCoO$_4$ (LSCO) and AFM LaSrMnO$_4$ (LSMO) to study the exchange bias effect in SrLaCo$_{0.5}$Mn$_{0.5}$O$_4$ compound. The dotted lines represent the interface. (a) LSCO and LSMO are repeated along the z-axis, (b & c) shows the interface with compensated and uncompensated AFM at the interface layers respectively, in which LSCO and LSMO are repeated along x- or y-axes to make the supercell. The exchange bias energy (EB) for each of the case is calculated by the spin-flip mechanism of the Co atoms in the FM LSCO layers. The spin-flip for Co atoms for each case are represented using the blue arrows. The results are obtained from DFT+U calculations with U = 5 eV. (d) For interface (a), (a-I) and (a-II) show the total and partial densities of states (DOS) before and after spin-flip respectively. Similarly, (b-I, b-II) and (c-I, c-II) show the total and partial DOS for interfaces (b) and (c) respectively.**

Under the collinear arrangement and the spin-flip mechanism, the EB is calculated as follows. In each of the supercell, the spin of the Co atoms in the FM LSCO layers are flipped, whereas, that of Mn atoms in AFM LSMO layers remain same. The exchange bias energy (E$_{EB}$) is the difference ($|E_I - E_{II}|$) between the two cases, i.e., between the spin-up (E$_I$) and spin-down (E$_{II}$) arrangement for Co atoms in the FM LSCO layer. A similar method has also been adopted earlier to study the exchange bias effect in SrRuO$_3$/SrMnO$_3$ [60]. In the first case (Fig. 6(a)), the EB is zero, suggesting weak magnetic coupling among the Co and Mn layers of atoms along the z-axis. This is due to the large layer separation of 6.31 Å.

In the second case, two possible types of interfaces are considered depending on the spin alignment of the Mn atoms (Fig. 6(b) and 6(c)). In these two types of interfaces, the spin alignment of the Mn atom at the interface layer is different. In the first case (Fig. 6(b)), the Mn spins at the interface layer are opposite to each other to form a compensated AFM structure. In the second structure (Fig. 6(c)), the spin-alignment of the Mn atoms in the interface layer are in the same direction to produce uncompensated moments at the interface. In case of the supercells having compensated and uncompensated LSMO AFM layers at the interface, the EBs are found to be 178.99 *meV/f.u.* and 82.33 *meV/f.u.* respectively. The Mn-Co distance in such supercells is 3.84Å. Compensated AFM layer at the interface, in principle, should not show any exchange bias effect [38]. However, in our calculation, for such an interface, we found there is a significant value of the exchange bias energy.

**Origin of exchange bias:** To explain the origin of the exchange bias energy, we shall consider the difference between the magnetic interaction energy ($J_{ij}$), given by the Hamiltonian,

$$H = -\sum_{i,j} J_{ij} \boldsymbol{S_i} \cdot \boldsymbol{S_j} \text{-----} (6)$$

First, we consider the case of the uncompensated LSMO AFM at the interface layer (Fig. 6 (c)) with the configuration I. The total magnetic interaction energy ($J_{tot}$) for this configuration is *$8J_1$-$8J_2$+$8J_3$*, where $J_1$, $J_2$, and $J_3$ are the interaction energies between Co-Co, Mn-Mn, and Co-Mn respectively. The interaction is considered positive for the same spin, whereas, they are considered to be negative for AFM ordering between two atoms. Similarly, for the configuration II, the total magnetic exchange interaction energy is *$8J_1$- $8J_2$-$8J_3$*. Therefore, the difference in magnetic energy between the two systems is 16$J_3$, which further leads to exchange bias energy.

To further analyze, in the lower panels of Fig. 6, we have plotted the total and Mn/Co-*d* DOS before and after spin-flip for the aforementioned interfaces. From the total DOS for each of these interfaces, we observe finite DOS at $E_F$ in either of the spin-channels, suggesting the FM metallic behavior. From the partial DOS plots, we find that Mn-*d* states create (pseudo) gap at $E_F$, which is similar to the pure AFM and insulating LSMO bulk compounds,[55] whereas, the Co-*d* states crosses the $E_F$ which is the case for intermediate spin state as observed earlier in FM and metallic LSCO compound [54]. This further confirms that the bulk magnetic phases are nearly maintained. Minor deviations are due to coupling between the Mn and Co spins across the interface.

For the interface along [001], we find that the DOS does not change with spin-flipping. This is because the strength of the magnetic coupling across the interface remained unaltered leading to the absence of an exchange bias effect for this interface. However, for the interface along $x$ and $y$, the DOS at $E_F$ changes significantly with spin-flipping suggesting a variation in the magnetic coupling. Therefore, the total energy of these configurations with spin-flipping changes to create an exchange bias effect.

## 5. Conclusions

**To summarize and conclude we have carried out a combined experimental and theoretical investigation to show that SLCMO produces an exchange bias as large as 5.5 kOe. To our understanding, this is the highest ever reported among all the transition metal layered oxides. The glassy magnetic nature of the sample has been confirmed using ac and dc magnetic measurements. As the first principles electronic calculations suggest the origin of this glassy phase and subsequently, the exchange bias effect is subscribed to the presence of competing magnetic interaction at the interface between magnetic domains at the sub-nanoscale. This work concludes that new layered oxides with more than one transition metals can be designed to create natural/artificial magnetic interfaces so that tunable giant exchange bias can be observed at the desired temperature.**

**Acknowledgments:**

**P.N.S. acknowledges the Council of Scientific and Industrial Research, India for financial support (Project No. 03 (1214)/12/EMR-II). We acknowledge Nano Functional Materials Technology Centre of Indian Institute of Technology Madras for X-ray Photoelectron spectroscopy measurement. We wish to thank Dr. A. Senyshyn for his help in the neutron diffraction experiment. The author T.C. gratefully acknowledges the financial support provided by FRM II to perform neutron scattering experiments at the Heinz Maier-Lieibnitz Zentrum (MLZ), Garching, Germany. The theoretical part of this work is**

**supported by DST, India through grant no. EMR/2016/003791. P.P. and B.R.K.N. would like to acknowledge the High Performance Computing Environment facility of Indian Institute of Technology Madras. P.P. acknowledges Indian Institute of Technology Madras for the Institute Postdoctoral fellowship. R.R.D. is grateful to Dr. P. Neenu Lekshmi for many fruitful discussions. R.R.D. thanks Indian Institute of Technology Madras Alumni for the travel support.**

**References:**

[1]  S. Baidya and T. Saha-Dasgupta, Physical Review B **84**, 035131 (2011).
[2]  J. Su, Z. Z. Yang, X. M. Lu, J. T. Zhang, L. Gu, C. J. Lu, Q. C. Li, J. M. Liu, and J. S. Zhu, ACS Appl Mater Interfaces **7**, 13260 (2015).
[3]  S. Yáñez-Vilar *et al.*, Physical Review B **84**, 134427 (2011).
[4]  N. Lee, H. Y. Choi, Y. J. Jo, M. S. Seo, S. Y. Park, and Y. J. Choi, Applied Physics Letters **104**, 112907 (2014).
[5]  R. von Helmolt, J. Wecker, B. Holzapfel, L. Schultz, and K. Samwer, Physical Review Letters **71**, 2331 (1993).
[6]  S. Jin, T. H. Tiefel, M. McCormack, R. A. Fastnacht, R. Ramesh, and L. H. Chen, Science **264**, 413 (1994).
[7]  Y. Tokura, Reports on Progress in Physics **69**, 797 (2006).
[8]  D. G. Schlom, L.-Q. Chen, X. Pan, A. Schmehl, and M. A. Zurbuchen, Journal of the American Ceramic Society **91**, 2429 (2008).
[9]  H. Y. Hwang, Y. Iwasa, M. Kawasaki, B. Keimer, N. Nagaosa, and Y. Tokura, Nat Mater **11**, 103 (2012).
[10] J. Chakhalian, A. J. Millis, and J. Rondinelli, Nat Mater **11**, 92 (2012).
[11] S. N. Ruddlesden and P. Popper, Acta Crystallographica **10**, 538 (1957).
[12] J. B. Torrance, Y. Tokura, A. I. Nazzal, A. Bezinge, T. C. Huang, and S. S. Parkin, Phys Rev Lett **61**, 1127 (1988).
[13] R. J. Cava, R. B. van Dover, B. Batlogg, and E. A. Rietman, Phys Rev Lett **58**, 408 (1987).
[14] S. B. Wilkins, P. D. Spencer, P. D. Hatton, S. P. Collins, M. D. Roper, D. Prabhakaran, and A. T. Boothroyd, Phys Rev Lett **91**, 167205 (2003).
[15] T. Birol, N. A. Benedek, and C. J. Fennie, Phys Rev Lett **107**, 257602 (2011).
[16] P D Battle *et al* , J. Phys.: Condens. Matter **8** L427 (1996).
[17] M. Lu, X. Deng, J. C. Waerenborgh, X. Wu, and J. Meng, Dalton Trans **41**, 11507 (2012).
[18] F. Tonus, C. Greaves, H. El Shinawi, T. Hansen, O. Hernandez, P. D. Battle, and M. Bahout, Journal of Materials Chemistry **21**, 7111 (2011).
[19] C. J. Zhang and H. Oyanagi, Physical Review B **79**, 064521 (2009).
[20] R. Ang, Y. P. Sun, X. Luo, C. Y. Hao, X. B. Zhu, and W. H. Song, Journal of Applied Physics **104**, 023914 (2008).
[21] T. Chakraborty, C. Meneghini, A. Nag, and S. Ray, J. Mater. Chem. C **3**, 8127 (2015).
[22] L. Xie and H. G. Zhang, Journal of Applied Physics **113**, 204506 (2013).
[23] R. C. Sahoo, S. K. Giri, P. Dasgupta, A. Poddar, and T. K. Nath, Journal of Alloys and Compounds **658**, 1003 (2016).

[24]  M. Y. Lin, Y. F. Wang, D. C. Ling, H. S. Sheu, and H. C. I. Kao, Journal of Superconductivity and Novel Magnetism **23**, 721 (2010).
[25]  P. Giannozzi *et al.*, J Phys Condens Matter **21**, 395502 (2009).
[26]  A. Cammarata and J. M. Rondinelli, Physical Review B **92**, 014102 (2015).
[27]  M. C. Biesinger, B. P. Payne, A. P. Grosvenor, L. W. M. Lau, A. R. Gerson, and R. S. C. Smart, Applied Surface Science **257**, 2717 (2011).
[28]  R. C. Sahoo, D. Paladhi, P. Dasgupta, A. Poddar, R. Singh, A. Das, and T. K. Nath, Journal of Magnetism and Magnetic Materials **428**, 86 (2017).
[29]  C. L. Bull, H. Y. Playford, K. S. Knight, G. B. G. Stenning, and M. G. Tucker, Physical Review B **94**, 014102 (2016).
[30]  Y. Sun, M. B. Salamon, K. Garnier, and R. S. Averback, Phys Rev Lett **91**, 167206 (2003).
[31]  J. A. Mydosh, Rep Prog Phys **78**, 052501 (2015).
[32]  J. A. Mydosh, *Spin glasses: an experimental introduction* (Taylor and Francis, 1993).
[33]  F. Wang, J. Zhang, Y.-f. Chen, G.-j. Wang, J.-r. Sun, S.-y. Zhang, and B.-g. Shen, Physical Review B **69**, 094424 (2004).
[34]  V. K. Anand, D. T. Adroja, and A. D. Hillier, Physical Review B **85**, 014418 (2012).
[35]  J. Souletie and J. L. Tholence, Physical Review B **32**, 516 (1985).
[36]  S. Sabyasachi, M. Patra, S. Majumdar, S. Giri, S. Das, V. S. Amaral, O. Iglesias, W. Borghols, and T. Chatterji, Physical Review B **86**, 104416 (2012).
[37]  A. D. Christianson, M. D. Lumsden, M. Angst, Z. Yamani, W. Tian, R. Jin, E. A. Payzant, S. E. Nagler, B. C. Sales, and D. Mandrus, Physical Review Letters **100**, 107601 (2008).
[38]  J. Nogués and I. K. Schuller, Journal of Magnetism and Magnetic Materials **192**, 203 (1999).
[39]  Manh-Huong Phan, Javier Alonso, Hafsa Khurshid, Paula Lampen-Kelley, Sayan Chandra, Kristen Stojak Repa, Zohreh Nemati, Raja Das, Óscar Iglesias and Hariharan Srikanth, Nanomaterials **6,** 221 (2016).
[40]  L. T. Coutrim *et al.*, Physical Review B **93**, 174406 (2016).
[41]  S. K. Giri, R. C. Sahoo, P. Dasgupta, A. Poddar, and T. K. Nath, Journal of Physics D: Applied Physics **49**, 165002 (2016).
[42]  Y.-k. Tang, Y. Sun, and Z.-h. Cheng, Physical Review B **73** (2006).
[43]  L. Klein, Applied Physics Letters **89**, 036101 (2006).
[44]  J. Geshev, Journal of Magnetism and Magnetic Materials **320**, 600 (2008).
[45]  M. I. B. S.V. Andreev, V.I. Pushkarsky, V.N. Maltsev, L.A. Pamyatnykha, and N. V. K. E.N. Tarasov, T. Goto, Journal of Alloys and Compour, ds **260**, 196 (1997).
[46]  M. Patra, M. Thakur, K. De, S. Majumdar, and S. Giri, J Phys Condens Matter **21**, 078002 (2009).
[47]  C. Binek, Physical Review B **70**, 014421 (2004).
[48]  S. K. Mishra, F. Radu, H. A. Durr, and W. Eberhardt, Phys Rev Lett **102**, 177208 (2009).
[49]  D. Senff, P. Reutler, M. Braden, O. Friedt, D. Bruns, A. Cousson, F. Bourée, M. Merz, B. Büchner, and A. Revcolevschi, Physical Review B **71**, 024425 (2005).
[50]  M. Merz, G. Roth, P. Reutler, B. Büchner, D. Arena, J. Dvorak, Y. U. Idzerda, S. Tokumitsu, and S. Schuppler, Physical Review B **74**, 184414 (2006).
[51]  J. Herrero-Martín, J. García, G. Subías, J. Blasco, and M. C. Sánchez, Physical Review B **72**, 085106 (2005).
[52]  M. Cwik, M. Benomar, T. Finger, Y. Sidis, D. Senff, M. Reuther, T. Lorenz, and M. Braden, Physical Review Letters **102**, 057201 (2009).
[53]  A. V. Chichev *et al.*, Physical Review B **74**, 134414 (2006).
[54]  H. Wu, Physical Review B **81**, 115127 (2010).

[55]     K. T. Park, Journal of Physics: Condensed Matter **13**, 9231 (2001).
[56]     C. Baumann, G. Allodi, A. Amato, B. Büchner, D. Cattani, R. De Renzi, R. Klingeler, P. Reutler, and A. Revcolevschi, Physica B: Condensed Matter **374-375**, 83 (2006).
[57]     C. Baumann, G. Allodi, B. Büchner, R. De Renzi, P. Reutler and A. Revcolevschi, Physica B: Condensed Matter **326**, 505 (2003).
[58]     S. Larochelle, A. Mehta, L. Lu, P. K. Mang, O. P. Vajk, N. Kaneko, J. W. Lynn, L. Zhou, and M. Greven, Physical Review B **71**, 024435 (2005).
[59]     H. Guo, Z. Hu, T.-W. Pi, H. L. Tjeng, and C. A. Komarek, Crystals **6,** 98 (2016).
[60]     S. Dong, Q. Zhang, S. Yunoki, J. M. Liu, and E. Dagotto, Physical Review B **84**, 224437 (2011).